\documentstyle[pra,preprint,aps]{revtex}
\tighten

\begin{document}
\title{Quantum History cannot be Copied}

\author{Arun K. Pati \footnote{Email: akpati@iopb.res.in} }
\address{Institute of Physics, Bhubaneswar-751005, Orissa, India }

\date{\today}
\maketitle
\def\ra{\rangle}
\def\la{\langle}
\def\ver{\arrowvert}

\begin{abstract}
We show that unitarity does not allow cloning of any two points in a ray.
This has implication for cloning of the geometric phase information in a 
quantum state. In particular, 
the quantum history which is encoded in the geometric phase during 
cyclic evolution of a quantum system cannot be copied. We also prove that 
the generalized geometric phase information cannot be copied by a
unitary operation. We argue that our result also holds in the consistent 
history formulation of quantum mechanics.

\end{abstract}

\vskip .5cm



\vskip 1cm

In quantum theory state of a single quantum is represented by not just
a vector $|\psi\ra$ in a separable Hilbert space ${\cal H}$, but by a
ray in the ray space ${\cal R}$. A ray
is a set of equivalence classes of states that differ from each other by
complex numbers of unit modulus. Thus the ray space ${\cal R}$ is defined as
${\cal R} =
\{ |\psi'\ra : |\psi \ra \sim |\psi'\ra = c|\psi \ra \}$,
where $c \in  {\bf C} $ is a group of non-zero complex numbers and $|c|=1$.
Given a quantum state $|\psi\ra$ we can generate a ray by the application
of the `ray operator' $R(c) = \exp[ i {\rm Arg}~( c ) |\psi\ra \la \psi| ]
= I + (c -1) |\psi\ra \la \psi|$ such that $R(c)|\psi \ra = |\psi' \ra$.
Geometrically, we can represent all these equivalent classes of states
as points in a ray and all of them represent the same physical state.
The set of rays of the Hilbert space ${\cal H}$ is called the projective 
Hilbert space ${\cal P} = {\cal H}/U(1)$. If we have a continuous unitary 
time evolution  of a quantum system $|\psi\ra \rightarrow U(t)|\psi \ra$ 
then the evolution can be represented as an open curve $\Gamma: 
t \rightarrow |\psi(t)\ra$ in ${\cal R}$  whose projection in ${\cal P}$ 
is also an open curve ${\widehat \Gamma}$. The quantum state at different 
times can belong to different rays. If the quantum state at two different times
belongs to the same ray, then it may trace an open curve $C$ in 
${\cal R}$, but its projection in ${\cal P}$ is a closed curve ${\widehat C}$.
Such an evolution is called a cyclic evolution.

In quantum information theory we view a quantum state $|\psi\ra$ as the
carrier
of both classical and quantum information. The fundamental unit of classical
information is a bit and that of quantum information is a qubit. Classical
bit can be copied but quantum bit or qubit cannot be copied. It is
the linearity
of quantum theory that does not allow us to produce a copy of an arbitrary
quantum state \cite{wz,dd}. Using unitarity one can prove that two
non-orthogonal states
cannot be copied either \cite{hpy}. However, orthogonal quantum states
like $|0\ra$ and
$|1\ra$ can be copied unitarily. Also, if we know a quantum state, we can make
as many copies as we wish. The subject of quantum cloning is an active area
of research. Recent studies have shown that one could make approximate copies
by deterministic transformations \cite{bh} and exact copies by 
probabilistic operations \cite{dg,akp0}. Another important limitation on 
quantum information is that it is impossible to delete an unknown quantum 
state \cite{pb}. We know that classical bit or quantum bit in orthogonal 
states can be deleted against a 
copy. However, linearity of quantum theory prohibits deletion of a qubit
against a copy.

Here, we ask the following question: If we can
make a copy of $|\psi\ra$ can we make copy of an equivalent state $|\psi'\ra$ 
by the same machine? That is whether $|\psi \ra \otimes |\Sigma\ra 
\rightarrow |\psi\ra \otimes |\psi \ra$ and $|\psi' \ra \otimes |\Sigma\ra 
\rightarrow |\psi' \ra \otimes |\psi' \ra$
is possible by a single unitary operator, where $|\Sigma\ra$ is the 
blank state and $|\psi'\ra = c|\psi\ra$.
Intuitively, one would say that
since $|\psi \ra$  and $ c|\psi\ra$ represent the same physical state from
informational point of view it should be possible to make copies of
$|\psi \ra$  and $ c|\psi\ra$ by the same cloning machine. But surprisingly,
this intuition is not correct. The proof is simple but nevertheless it is
important. This has important implication in cloning of relative phase 
information. For example, we prove that the geometric phase information during
cyclic evolution of a quantum system cannot be copied by unitary machine or 
any physical operation (a completely positive trace preserving map).
Furthermore, we prove that the non-cyclic geometric phase cannot be 
copied during an arbitrary evolution of a quantum system. The 
important implication of our theorems is that history of a quantum system 
which is encoded in the geometric phase cannot be copied. 
Interestingly, we argue that our result also holds in the  consistent 
histories formulation of quantum theory where the geometric phase appears 
naturally in the histories.
\\

\noindent
{\bf Theorem 1:} {\it In general, two points in a ray cannot be cloned by a 
single unitary machine.}\\

\noindent
{\it Proof:} A ray is an equivalence classes of states up to 
global phases . Thus, rather than the usual cloning machine having the 
form $|\psi \rangle \otimes |\Sigma \ra \rightarrow  |\psi \ra \otimes 
|\psi \ra$ we should really allow for
\begin{equation}
|\psi\ra \otimes |\Sigma \ra  \rightarrow  
e^{i \theta(\psi) } |\psi \ra \otimes |\psi \ra,
\end{equation}
where $\theta(\psi)$ can be an arbitrary function of $|\psi \ra$.
Now, for two equivalent states $|\psi \ra$ and $|\psi' \ra$
if we take the transformation 
\begin{eqnarray}
|\psi \rangle \otimes |\Sigma \ra &\rightarrow &  
|\psi \ra \otimes |\psi \ra\nonumber\\
|\psi' \rangle \otimes |\Sigma \ra &\rightarrow & |\psi'\ra \otimes 
|\psi' \ra, 
\end{eqnarray}
then this says that which specific member of the 
equivalence class we are in that needs to be preserved. One might argue 
that this would 
be against the spirit of having an equivalence class in the first place 
where any member of the equivalence class may be substituted with any 
other at any time. So, the general cloning transformation may be given by (1).
Note that this ambiguity does not arise in usual cloning literature 
that describe approximate clones because there one asks only about the 
reduced state of each subsystem. It should be mentioned that
we can also use the transformation (2) to prove our theorem but (1) is
more general way of stating the action of cloning map on a ket. 
As we know for a fixed state vector, an overall phase is not important, 
since physical states are density matrices and not vectors. 
Indeed the cloning maps given by (1) and (2) are equivalent. Unless there is 
confusion, in this paper, we use sometime transformations (1) or (2).

Let us consider two equivalent classes of states $|\psi \ra$ and 
$|\psi' \ra$. The cloning transformation 
for two points in a ray is now given by 
\begin{eqnarray}
|\psi \ra \otimes |\Sigma \ra  &\rightarrow&  
e^{i \theta(\psi) } |\psi \ra \otimes |\psi \ra \nonumber\\
|\psi' \ra \otimes |\Sigma \ra  &\rightarrow&  
e^{i \theta(\psi') } |\psi \ra \otimes |\psi \ra
\end{eqnarray}
Since unitarity preserves the inner product we must have 
$c = \exp[ i \theta(\psi) - \theta(\psi')]$. However, this cannot hold for 
arbitrary values of $c, \theta(\psi)$ and $\theta(\psi')$. This proves 
that two  equivalent states cannot be copied by the same machine even 
if we know the state. The physical meaning of this theorem is that the
relative phase between two points in a ray cannot be copied by a unitary
machine.

The `no-cloning theorem for a ray' will have implication for cloning of 
relative phase information in quantum systems. In recent years, there 
have been 
considerable interest in the study of the relative phases and in particular 
the Berry phases \cite{berry} in 
quantum systems. It is also hoped that the Berry phase which is of geometric 
origin may be used in design of robust logic gates in quantum computation. The
original Berry phase was
discovered in the context of quantum adiabatic theorem \cite{lis,mes}.
It is basically an extra phase
that the system acquires when the Hamiltonian is slowly changed
cyclically over one time period. The Berry phase is independent of the 
detailed dynamics of the system and is of purely geometric in origin. 
This arises due to the non-trivial curvature of the parameter space in 
which the state vector is transported 
around a closed loop. An early discovery of the geometric phase was made by 
Pancharatnam in the context of interference of light \cite{panch}. 
The Berry phase was then generalized to non-adiabatic
but cyclic evolutions of quantum system by Aharonov and Anandan \cite{aa}.
In fact, now we know that the geometric phase appears in much more general
context than it was thought before \cite{samu,ms,akp}.

Consider the unitary time
evolution of a quantum system where the state vector evolves as $|\psi(0)\ra
 \rightarrow  |\psi(t) \ra = U(t)|\psi(0) \ra$ such that at $t=T$,  
$|\psi(T) \ra = e^{i\Phi}
|\psi(0) \ra$, $\Phi$ being the total phase. That is to say that the quantum 
system at $t=T$ comes back to its original state
apart from a phase factor. As we know such an evolution is called the 
cyclic evolution. 
Even though $|\psi(0) \ra$ and $|\psi(T) \ra$ are equivalent it is the relative
phase $\Phi$ between them that is observable.
The total phase $\Phi$ that the system acquires during a cyclic evolution is
composed of two phases, one is the dynamical phase $\delta$ and the
other is the geometric phase $\beta({\widehat C})$. 
This $\beta({\widehat C})$ is 
also known as the Aharonov-Anandan (AA) phase \cite{aa}.
Thus the total phase is given by
\begin{equation}
\Phi = \delta + \beta({\widehat C}),
\end{equation}
 where $\delta$ is the dynamical phase and $\beta({\widehat C})$ is 
the geometric phase. The dynamical phase is given by 
\begin{equation}
 \delta = - \frac{1}{\hbar} \int_0^T \la \psi(t)|H|\psi(t)\ra ~dt.
\end{equation}
It represents, in a sense, an `internal clock' of the quantum system. On the 
other hand the geometric phase is given by 
\begin{equation}
\beta({\widehat C}) = i\oint \la {\tilde \psi}(t)|{\dot {\tilde \psi}}(t) 
\ra ~dt = i\oint \la {\tilde \psi}|d {\tilde \psi} \ra,
\end{equation}
where $|{\tilde \psi}(t) \ra = \exp(-if(t)) |\psi(t)\ra$ with $f(t)$ being any
smooth function that satisfies $f(T)- f(0)= \Phi$.
Here, $ i \la {\tilde \psi}|d {\tilde \psi} \ra$ is the differential
connection-form that gives rise to geometric phase. It is gauge invariant,
reparameterization invariant, and depends only on the closed curve 
${\widehat C}$ in the projective Hilbert space ${\cal P}$ of the quantum 
system. Unlike the
dynamical phase, the geometric phase indeed depends on the path of the 
evolution \cite{pati91} and is a non-integrable quantity.

In this context, if we are able to clone $|\psi(0) \ra$, one may think
that we can also clone
$|\psi(T) \ra$ as they really belong to the same ray. This is because during
a cyclic evolution the system starts from a ray and after a time period $T$
comes back to the same ray but at a different point. Now, application of our
theorem tells us that we cannot clone $|\psi(0) \ra$ and $|\psi(T) \ra$
by a single unitary machine. \\

\noindent
{\bf Proposition:} {\it Quantum history which is encoded in the geometric phase
during cyclic evolution of a quantum system cannot be copied by a unitary 
transformation.}\\

\noindent
{\it Proof:} 
Suppose we could copy $|\psi(0) \ra$ and $|\psi(T) \ra$
by a unitary machine. That is
\begin{eqnarray}
|\psi(0) \ra \otimes |\Sigma\ra & \rightarrow & |\psi(0) \ra \otimes 
|\psi(0) \ra \nonumber\\
|\psi(T) \ra \otimes |\Sigma\ra & \rightarrow & |\psi(T) \ra \otimes 
|\psi(T) \ra.
\end{eqnarray}
By unitarity, Eq(7) implies that we must have $\la \psi(0)|\psi(T)\ra 
\rightarrow \la \psi(0)|\psi(T)\ra^2$, i.e., $\exp(i\Phi) \rightarrow 
 \exp(2 i\Phi)$. However, this is not possible by a unitary machine. 
If it is possible, then
that would mean $\exp(i\Phi) =1$ which is not true in general. Moreover, 
if the system undergoes parallel transportation then it
acquires a pure geometric phase and we will have $\Phi = \beta({\widehat C})$
\cite{aa}.
Here we would like to mention that an {\em arbitrary quantum
state cannot be parallel transported} either. (This is another no-go 
theorem. See the notes in the end). 
Then the no-cloning theorem for a ray tells us that we cannot clone
the geometric phase
information  of a quantum system during a cyclic evolution. Since the 
geometric phase attributes memory
to a quantum system it remembers the history of the evolution. This then 
implies that the {\em quantum history cannot be copied.} Therefore, even 
if we can make copy of a known quantum states we cannot copy its history. 
The only way to copy the history is to first make a copy of the state and 
then pass the copied quantum system through the same cycle again. 
Physical reason for this
impossibility is the following. We are able to make copy of $|\psi(0) \ra$
because we have complete knowledge of it. But the geometric phase not only
depends on $|\psi(0) \ra$ but also on the path of the evolution that the
system has undergone in the past. Unless we have knowledge of the past
history we cannot copy the geometric phase. 

In fact, we can prove that quantum history cannot be copied by any physical
operation. A physical operation in quantum theory we mean a completely 
positive (CP) trace preserving mappin g.\\

\noindent
{\bf Theorem 2:} {\it In general, geometric phase during a cyclic evolution 
cannot be copied by a completely positive map.}\\

\noindent
{\it Proof:} We know that by including ancilla, any CP map can be realized 
as a unitary evolution in an enlarged Hilbert space. Consider the cloning of
$|\psi(0) \ra$ and $|\psi(T) \ra$ including ancilla. This may be given by
\begin{eqnarray}
|\psi(0) \ra \otimes |\Sigma \ra \otimes | A \ra & \rightarrow & 
|\psi(0) \ra \otimes |\psi(0) \ra \otimes | A(0) \ra \nonumber\\
|\psi(T) \ra \otimes |\Sigma \ra \otimes  |A \ra & \rightarrow & 
|\psi(T) \ra \otimes |\psi(T) \ra \otimes |A(T) \ra,
\end{eqnarray}
where $|A \ra$ is the initial state and $|A(0)\ra$, $|A(T)\ra$ 
are the final states of the ancilla. 
By unitarity in the enlarged Hilbert space, we have 
\begin{eqnarray}
e^{i\Phi} = e^{2i\Phi} \la A(0)|A(T)\ra.
\end{eqnarray}
This cannot hold in general, hence it is impossible to copy quantum history by
any physical operation. However, if it so happens that environment also 
undergoes a cyclic evolution and acquires equal and opposite relative phase
as that of the quantum system, i.e.,  $|A(T)\ra = \exp(-i \Phi) |A(0) \ra$, 
then possibly quantum history can be copied by a completely positive map. 

Next, we prove that the geometric phase information during a general quantum 
evolution cannot be copied. When a quantum system evolves in time it
traces an open path $\Gamma: t \rightarrow |\psi(t)\ra$ in the ray space 
${\cal R}$ and the quantum state at 
different times belong to different rays. The projection of the evolution 
path $\Gamma$ in the projective Hilbert space ${\cal P}$ is also an open 
path ${\widehat \Gamma}$. 
For example, consider the time evolution
$|\psi(0)\ra \rightarrow |\psi(t)\ra$. The evolution under 
consideration {\em need not be adiabatic, cyclic, and even unitary}. 
All that is required is that there is a linear map and evolution curve 
should be smooth with a inner product defined over the Hilbert space.
In this case the initial and final states are not equivalent. For any two 
non-orthogonal states the relative 
phase (or total phase difference) between 
them is given by the Pancharatnam phase $\Phi_P = 
{\rm Arg} \la \psi(0)|\psi(t) \ra$ \cite{panch,samu}. 
It is always possible to write the total phase $\Phi_P$ as sum of the 
dynamical phase $\Phi_D$ and the geometric phase $\Phi_G$. The dynamical 
phase is given by
\begin{equation}
 [\Phi_D]_0^t = - i \int \la \psi(t)|{\dot \psi}(t)\ra ~dt.
\end{equation}
It depends on the detailed dynamics that the system is undergoing.
The geometric phase is given by 
\begin{eqnarray}
[\Phi_G]_0^t & = & 
{\rm Arg} \la \psi(0)|\psi(t) \ra + 
i \int \la \psi(t)|{\dot \psi}(t)\ra ~dt\nonumber\\ 
& = & i\int \la \chi(t)|{\dot \chi}(t) \ra ~dt
= i\int \la \chi|d \chi \ra,
\end{eqnarray}
where $|\chi(t) \ra = \frac{ \la \psi(t)| \psi(0)\ra}
{ |\la \psi(t)| \psi(0)\ra| } |\psi(t) \ra$ is a reference-section 
introduced in \cite{akp,akp1}.
Here, $ i \la \chi|d \chi \ra$ is the differential
connection-form that gives rise to the most general geometric phase. 
It is again $U(1)$ gauge invariant, reparameterization invariant, 
and depends only on the geometry of the open curve ${\widehat \Gamma}$ 
in the projective 
Hilbert space ${\cal P}$ of the quantum system. It can be
shown that it is a {\em non-additive quantity} which in turn implies 
that the system 
remembers along which path it has been transported. Thus, the most general 
geometric phase remembers the history of quantum system and attributes a 
memory to the quantum system. Next we prove the following.\\

\noindent
{\bf Theorem 3:} {\it Quantum history which is encoded in the generalized 
geometric phase during arbitrary evolution of a quantum system
cannot be copied unitarily.}\\

\noindent
{\it Proof:} Consider a sequence of time evolution of a quantum system 
from $t=0$ to $t_1$ and then from $t_1$ to $t_2$. Thus, we have the time 
evolution $|\psi(0)\ra \rightarrow |\psi(t_1)\ra \rightarrow |\psi(t_2)\ra$. 
Suppose, we want to clone the quantum states $|\psi(0)\ra$, $|\psi(t_1)\ra$
and $|\psi(t_2)\ra$ by a unitary machine. Then we will have the following
cloning transformation
\begin{eqnarray}
|\psi(0) \ra \otimes |\Sigma\ra & \rightarrow & |\psi(0) \ra \otimes 
|\psi(0) \ra \nonumber\\
|\psi(t_1) \ra \otimes |\Sigma\ra & \rightarrow & |\psi(t_1) \ra \otimes 
|\psi(t_1) \ra \nonumber\\
|\psi(t_2) \ra \otimes |\Sigma\ra & \rightarrow & |\psi(t_2) \ra \otimes 
|\psi(t_2) \ra.
\end{eqnarray}
Now, by unitarity, taking all the inter inner products we will have 
\begin{eqnarray}
\la \psi(0)|\psi(t_1) \ra \la \psi(t_1)|\psi(t_2) \ra \la \psi(t_2)
|\psi(0) \ra = 
[\la \psi(0)|\psi(t_1) \ra \la \psi(t_1)|\psi(t_2) \ra \la \psi(t_2)
|\psi(0) \ra]^2.
\end{eqnarray}
Let us define a complex quantity called as the three point Bargmann invariant 
$\Delta^{(3)}$ as
\begin{equation}
\Delta^{(3)} = \la \psi(0)|\psi(t_1) \ra \la \psi(t_1)|\psi(t_2) \ra 
\la \psi(t_2)|\psi(0) \ra.
\end{equation}
The three point Bargmann invariant remains the same under unitary and 
antiunitary transformation and plays an important role in the 
kinematic approach
to the theory of geometric phases developed by Mukunda and Simon \cite{ms}. 
Now taking the argument of both the sides of Eq(13), we will have 
${\rm Arg} \Delta^{(3)} \rightarrow 2 {\rm Arg} \Delta^{(3)}$ which implies 
that ${\rm Arg} \Delta^{(3)} = 0$. But
 ${\rm Arg} \Delta^{(3)}$ is nothing but the excess geometric phase that 
the system may acquire in going from $t_0 = 0$ to $t_1$ and then from $t_1$ 
to $t_2$ instead of going from $t_0 = 0$ to $t_2$ directly \cite{akp}.
More precisely, 
let $[\Phi_G]_{0}^{t_1}$ is the geometric phase that the system acquires 
during the evolution from time $t_0=0$ to $t_1$, $[\Phi_G]_{t_1}^{t_2}$
is the geometric phase that the system acquires 
during the evolution from time $t_1$ to $t_2$, and $[\Phi_G]_{0}^{t_2}$
is the geometric phase that the system acquires 
during the evolution directly from time $t_0=0$ to $t_2$. An important 
property of the geometric phase is that it is {\em non-additive}.
We will indeed have 
$ [\Phi_G]_{0}^{t_1} + [\Phi_G]_{t_1}^{t_2} \not= [\Phi_G]_{0}^{t_2}$.
Thus, the excess geometric phase given by
\begin{equation}
[\Phi_G]_{0}^{t_1} + [\Phi_G]_{t_1}^{t_2} - [\Phi_G]_{0}^{t_2} =
{\rm Arg} [ \la \psi(0)|\psi(t_1) \ra \la \psi(t_1)|\psi(t_2) \ra 
\la \psi(t_2)|\psi(0) \ra] = {\rm Arg} \Delta^{(3)}.
\end{equation}
In the above the dynamical phase has disappeared because that is simply an
additive quantity. Thus, ${\rm Arg} \Delta^{(3)} \not=0$. Hence this shows 
that the quantum history encoded in the generalized geometric phase during
an arbitrary quantum evolution cannot be copied.

Our result has implication in the context of consistent history formulation
of quantum mechanics developed by Griffiths \cite{grif}, Omne \cite{omn} and 
Gell-Mann and Hartle \cite{hart}. Recently, it has been shown by Anastopoulos 
and Savvidou
that the geometric phase is manifested in the probabilistic structure of 
histories. Specifically, they have shown that the geometric phase is 
the basic building block of the
interference phase between pair of histories \cite{ana}.

In consistent history approach, typically a history can be thought of as
a sequence of properties
or events which correspond to a time-ordered sequence of propositions about
the quantum system. These events are represented by projectors $\Pi_1, \Pi_2,
\ldots \Pi_n$ on the Hilbert space ${\cal H}$ at a succession of
times $t_1 < t_2 < \cdots t_n$. These projectors at different times need not
commute. One defines the space of all histories by a history Hilbert space 
${\cal H}_h = {\cal H}_{t_i}^{\otimes t_i}$ which consists of tensor 
product of copies of the Hilbert space ${\cal H}_{t_i}$ at $i$th instant
of time. On the history Hilbert space 
one represents history as a projector $P= \Pi_1 \otimes \Pi_2 \cdots 
\Pi_n$. The meaning of such a history is that events $\Pi_i$ occurs in the
system at time $t_i$, respectively. One assigns a realistic interpretation
to such a history provided certain consistency conditions are satisfied.

By introducing Heisenberg projector $ \Pi_i(t_i) = U(t_i)^{\dagger}
\Pi_i U(t_i)$ one can define a weight operator $C_{P}$ to each history
defined by 
\begin{equation}
C_{P} =  U(t_n)^{\dagger} \Pi_n U(t_n) \cdots U(t_1)^{\dagger}
\Pi_1 U(t_1) = \Pi_n(t_n) \cdots \Pi_1(t_1).
\end{equation}
Given a pair of weight operators $C_P$ and $C_{P'}$ for two histories 
one defines an important quantity --that is the decoherence functional 
which is a complex valued function $d(P, P')$ of history propositions $P, P'$
that measures the quantum mechanical interference between them. More 
precisely, the decoherence functional is a mapping 
$d: {\cal H}_h \times {\cal H}_h \rightarrow {\bf C}$ that satisfies
the following conditions \cite{isham}:\\
1. $d(P, P') = d(P', P)^*$ for all $P, P'$. This is hermiticity.\\
2. $d(P, P) \ge 0$ for all $P$. This is positivity.\\
3. If $P$ and $P'$ are orthogonal, then for all $P''$, 
$d(P \oplus P', P'') = d(P, P'') + d(P', P'')$. This is additivity.\\
4. $d(I, I) = 1$.  This is normalization.

One way of defining the decoherence functional is 
\begin{equation}
d(P, P') = {\rm Tr} \big( C_P \rho_0 C_{P'}^{\dagger} \big)
\end{equation}
for an initial density matrix $\rho_0$. This can be interpreted as 
a probability in standard quantum theory under
certain conditions. When $d(P, P') = 0$ for $P\not= P'$ in a set of histories
that satisfies $\sum_i \Pi_i = 1$ and $P P' = P \delta_{PP'}$, then under this
condition $d(P, P)$ can be regarded as the probability that the history 
proposition $P$ is true. The decoherence functional $d(P, P')$ can be thought 
of as the degree of interference between the histories $P$ and $P'$.

To see how the geometric phase appears in the consistent history formulation,
consider the time ordered events with projectors at different times as
$\Pi_0, \Pi_1, \ldots \Pi_n$. If one assumes that the projectors are fine
grained, they can be represented as elements of the projective Hilbert space
${\cal P}$, i.e., $\Pi_i= \Pi_{t_i} = |\psi(t_i)\ra \la \psi(t_i)|$. If one 
neglects the dynamical evolution, i.e., set the Hamiltonian equal to zero, 
then trace of the weight operator is given by
\begin{equation}
{\rm Tr} C_P =  \la \psi(0)|\psi(t_n) \ra \la \psi(t_n)|\psi(t_{n-1}) \ra 
\cdots \la \psi(t_1)|\psi(0) \ra.
\end{equation}
By assuming the number of time steps $n$ very large and each time steps
differs by $\delta t$ with $\delta t \sim O(1/n)$ one can approximate this
by a continuous time history and we have 
\begin{equation}
{\rm Arg}[ {\rm Tr} C_P] = \Phi_G.
\end{equation}
That is to each history one can assign a geometric phase \cite{ana}. 
Actually, one can understand the origin of the geometric phase in history 
formulation as follows. Note that 
\begin{equation}
{\rm Tr}C_P = {\Delta^{(n+1)}}^*,
\end{equation}
where $\Delta^{(n+1)}$ is the $(n+1)$-point Bargmann invariant defined as 
\begin{equation}
\Delta^{(n+1)}=  \la \psi(0)|\psi(t_1) \ra \la \psi(t_1)|\psi(t_2) \ra 
\cdots \la \psi(t_{n-1})|\psi(t_n) \ra.
\end{equation}
We already know that in the standard quantum theory the Bargmann invariant 
represents the excess geometric phase if system undergoes sequence of times
evolution between $t_0 < t_1 < \cdots t_n $ and a direct evolution from 
$t_0$ to $t_n$. This is true even if we do not assume that the Hamiltonian
is zero. So in the consistency history formulation the appearance of geometric 
phase is natural. It has been also shown that the information about the 
geometric phase for a set of histories is sufficient to reconstruct the 
decoherence functional \cite{ana}.
Now, from our theorem we know that geometric phase cannot be copied, 
then this also holds for the decoherence functional in the consistent 
history formulation. 

 In conclusion, we have proved that two equivalence classes of states
 representing the same physical state cannot be cloned by a unitary operator.
It is the relative phase between two points in a ray that is impossible to 
clone.
Even though the proof is really simple its implications may be important.
During cyclic evolution of a quantum system the initial and final states are
equivalent and can be represented as two points in a ray. 
One implication of our theorem is that the geometric phase information 
during a cyclic
evolution cannot be copied. Since the geometric phase attributes a memory 
to a quantum system and remembers the history, this suggests that even 
though a state can be copied its quantum history cannot be copied. 
We have also shown that the
geometric phase information during a cyclic evolution cannot be copied 
by a physical operation. In addition, we have proved that the geometric 
phase during arbitrary quantum evolution cannot be copied by a unitary machine.
Interestingly, we have argued that our result also holds in the consistent 
history formulation of quantum theory. We hope that the impossibility of
copying quantum history may have some application is quantum cosmology.

{\it Notes:} Here we prove that in general an arbitrary quantum state
cannot be parallel transported. The parallel transport condition for a
pure quantum state is
that it never rotates locally (so it does not acquire any phase
infinitesimally) but can under go a net rotation globally.
This reflects the curvature of the quantum state space, i.e., the projective 
Hilbert space ${\cal P}$ in which the
vector is parallel transported.
Thus, during a parallel transportation a state can acquire a phase
if brought back to its original position along a closed path ${\widehat C}$. 
This phase is essentially the holonomy angle or the geometric phase
$\beta({\widehat C})$. Mathematically, the parallel transport condition 
for a vector $|\psi(t) \ra$ can be expressed as 
$\la \psi(t)|{\dot \psi}(t)\ra =0$.
If $|\psi(t) \ra$ satisfies this then $|\psi(T) \ra = 
\exp(i\beta({\widehat C}) )
|\psi(0) \ra$ \cite{aa} during a cyclic evolution and 
$ \la \psi(0)|\psi(t) \ra =  |\la \psi(0)|\psi(t) \ra| \exp[\Phi_G]$
during arbitrary non-cyclic evolution \cite{akp,akp1}.

Suppose we have a set of known orthonormal bases $\{ |\psi_n(t)\ra \}$ with
$\la \psi_n(t)|\psi_m(t)\ra = \delta_{nm}$ for all $t$. If these bases 
undergo parallel transportation then they satisfy 
$\la \psi_n(t)|{\dot \psi_n}(t)\ra =0$. Now let $|\psi(t) \ra$ be
an arbitrary state: $|\psi(t) \ra = \sum_{n=1}^N c_n(t) |\psi_n(t) \ra \in
{\cal H}^N $. The question is if $\la \psi_n(t)|{\dot \psi_n}(t)\ra =0$
does that mean
$\la \psi(t)|{\dot \psi}(t)\ra =0$? The answer is no. To be explicit we have
\begin{equation}
 \la \psi(t)|{\dot \psi}(t)\ra = \sum_n c_n^* {\dot c}_n +
 \sum_{nm, n\not=m} c_m^* c_n \la \psi_m(t)|{\dot \psi_n}(t)\ra \not= 0.
\end{equation}
 Thus, in general an arbitrary quantum state cannot be parallel transported.
In other words, {\em universal parallel transportation machine cannot exist}.
However, there can be some special cases where it can be.
When $c_n(t)$'s are time-independent and
$\la \psi_m(t)|{\dot \psi_n}(t)\ra =0$
for $m\not=n$ then an arbitrary state can also undergo parallel
transportation. This is a both necessary and sufficient condition.
It would be an interesting problem by itself to find what kind of
Hamiltonian would satisfy the parallel transport condition for an arbitrary
quantum state.



\begin{thebibliography}{99}



\bibitem{wz} W. Wootters and W. H. Zurek, Nature (London), {\bf 299},
802 (1982)

\bibitem{dd} D. Dieks, Phys. Lett. A {\bf 92}, 271 (1982).

\bibitem{hpy} H. P. Yuen, Phys. Lett. A {\bf 113}, 405 (1986).

\bibitem{bh} V. Buzek and M. Hillery, Phys. Rev. A {\bf 54}, 1844 (1996).

\bibitem{dg} L. M. Duan and G. C. Guo, Phys. Rev. Lett. {\bf 80},
4999 (1998).

\bibitem{akp0} A. K. Pati, Phys. Rev. Lett. {\bf 83}, 2849 (1999).

\bibitem{pb} A. K. Pati and S. L. Braunstein, Nature, {\bf 404}, 164 (2000).

\bibitem{berry} M. V. Berry, Proc. Roy. Soc. (Lond.) {\bf 392}, 45 (1984).

\bibitem{panch} S. Pancharatnam, Proc. Indian Acad. Sci. A {\bf 44}, 
247 (1956).



\bibitem{lis} L. I. Schiff, {\it Quantum Mechanics} (McGraw-Hill, 1968).

\bibitem{mes} A. Messiah, {\it Quantum Mechanics} 
(North-Holland, Amsterdam, 1962).

\bibitem{aa} Y. Aharonov and J. Anandan,
Phys. Rev. Lett.{\bf 58}, 1539 (1987).

\bibitem{samu} J. Samuel and R. Bhandari, Phys. Rev. Lett. {\bf 60}, 
2339 (1988).

\bibitem{ms} N. Mukunda and R. Simon, Ann. Phys. {\bf 228}, 205 (1993).


\bibitem{akp} A. K. Pati, Phys. Rev. A {\bf 52}, 2576 (1995)

\bibitem{akp1} A. K. Pati, J. of Phys. A {\bf 28}, 2087 (1995).

\bibitem{pati91} A. K. Pati, Phys. Lett. A {\bf 159}, 105 (1991)


\bibitem{grif} R. B. Griffiths, J. Stat. Phys. {\bf 36}, 219 (1984).

\bibitem{omn} R. Omnes, J. Stat. Phys. {\bf 53}, 893 (1988).

\bibitem{hart} M. Gell-Mann and J. B. Hartle, Phys. Rev. D {\bf 47}, 
3345 (1993).

\bibitem{ana} C. Anastopoulos and N. Savvidou, Int. J. Theo. Phys. {\bf 41}, 
1572 (2002).

\bibitem{isham} C. J. Isham, Int. J. Theo. Phys. {\bf 36}, 785 (1997).

\end{thebibliography}
\end{document}